\documentclass[
    ,final            
  ]
  {aipproc}

\layoutstyle{8x11double}

\newcommand{\lsim}{\raise.35ex\hbox{$<$}\kern-0.75em\lower.5ex\hbox{$\sim$}}
\newcommand{\gsim}{\raise.35ex\hbox{$>$}\kern-0.75em\lower.5ex\hbox{$\sim$}}
\begin{document}

\title{Spin fluctuations and weak pseudogap behaviors in Na$_{0.35}$CoO$_2$:renomarization of band structure}

\classification{74.25.-q, 74.25.Ha, 74.25.Jb}
\keywords      {FLEX, Na$_x$CoO$_2$, spin fluctuations, pseudogap}

\author{Keiji Yada}{
  address={Department of Physics, Nagoya University,
 Nagoya, 464-8602, JPN}
}

\author{Hiroshi Kontani}{
  address={Department of Physics, Nagoya University,
 Nagoya, 464-8602, JPN}
}

\begin{abstract}
We analyze the normal electronic states of Na$_{0.35}$CoO$_2$
based on the multi-orbital Hubbard model using the FLEX approximation.
The fundamental electronic property of this system is drastically changed
by the presence or absence of the small hole pockets associated with the $e_g'$ orbital.
This change of the Fermi surface topology
may be caused by the crystalline electric splitting due to the trigonal distortion.
When small hole pockets are absent,
the weak pseudogap bahaviors appear in the density of states and the uniform spin susceptibility,
which are observed by recent experiments.
We estimate the mass enhancement factor of quasiparticle $m^*/m\simeq 1.5\sim 1.8$.
This result suppports ARPES measurements.
\end{abstract}

\maketitle

Na$_{0.35}$CoO$_2\cdot1.3$H$_2$O is the first Co-oxide supercondoctor with $T_c\sim 4.5K$\cite{Takada}.
While many theretical and experimental studies are widely performed,
the topology of Fermi surface (FS) and the low-energy
electronic structure are still unresolved.
To resolve these problems is very important to find out the mechanism of superconductivity.

In Na$_x$CoO$_2$, local density approximation (LDA)
calculations\cite{Singh} have predicted that Na$_x$CoO$_2$ has a large FS
associated with the $a_{1g}$ band and six small hole pockets
corresponding to the $e'_g$ band.
However, such small pockets are not observed in recent ARPES measurements\cite{ARPES1,ARPES2}.
ARPES measurements also observed that
renomarized quasiparticle bandwidth is approximately half of that calculated by band calculation.
In the present work, we study the normal electronic states in
Na$_{0.35}$CoO$_2$ using the fluctuation exchange (FLEX) approximation
to elucidate Fermi surface topology and renormalized band structure.
In Na$_x$CoO$_2$, the topology of the FS is sensitively changed
by the $a_{1g}$-$e'_g$ splitting $3V_t$, whose value can be modified by the trigonal distortion of crystal.
In this work, we study the many body effect for various values of $3V_t$.

We have reported\cite{we} that density of states (DOS) on the Fermi energy and uniform
spin susceptibility increase as temperature decreases when small
pockets exist.
On the other hand, when small pockets are absent, both of them decrease
at lower temperatures. It is a weak pseudogap behavior due to
magnetic fluctuations.
In this case, the degree of reduction of DOS is greater
when the top of the $e'_g$ band is just below the Fermi level.
In experimental measurements, both uniform spin susceptibility\cite{Sato-susc} and
DOS in the photoemission spectroscpy\cite{PES}
decrease at lower temperatures. These pseudogap behaviors are consistent
with the latter result.
As the pseudogap in the DOS is more prominent in bilayer hydrate samples
than that in monolayer ones, the effect of intercalation of
water is expected to raise the $e'_g$ band slightly
as expected by the analysis based on the point charge model.
As a result, we have concluded that the small Fermi pockets do not exist or very small if any.

In this paper, we report the band structure of Na$_{0.35}$CoO$_2$ at $3V_t=0.12$.
In this case, small hole pockets are absent; they sink just below the Fermi energy.
We find that the quasiparticle band is renormalized by the electronic correlation,
and mass enhancement factor $m^*/m\simeq 1.5\sim 1.8$.
It agrees with the result of ARPES measurements.

The model Hamiltonian used in the present study is as follows.
\begin{eqnarray*}
H&=&H_0+H',
\end{eqnarray*}
\begin{eqnarray*}
H_0&=&\sum_{i,j,\sigma}\sum_{\ell\ell'}t^{\ell\ell'}_{ij} \ c^\dag_{i\ell\sigma} c_{j\ell'\sigma} \\
   &=&\sum_{{\bf k},\sigma}\sum_{\ell\ell'}\varepsilon_{\bf k}^{\ell\ell'} \ c^\dag_{{\bf k}\ell\sigma} c_{{\bf k}\ell'\sigma},
\end{eqnarray*}
where $\ell$ represents $3d$ orbitals of Cobalt atoms (5 orbitals) and $2p$ orbitals of Oxygen atoms (2 $\times$ 3 orbitals),
and $\hat t$ has the Slater-Koster's matrix form.
$H'$ is on-site Coulomb integrals in $t_{2g}$($d_{xy},d_{yz},d_{zx}$) orbitals.
\begin{eqnarray*}
H'&=&H_U+H_{U'}+H_J+H_{J'},
\end{eqnarray*}
where $U$ ($U'$) is the intra-orbital (inter-orbital)
 Coulomb interaction,
$J$ is the Hund's coupling and $J'$ represents the pair-hopping interaction.
We put $U'=1.3$, $J=J'=0.13$, $U=U'+2J=1.56$ hereafter.
In these parameters, deformation of the FS due to the interaction is very small.
\begin{figure}
  \includegraphics[height=.25\textheight]{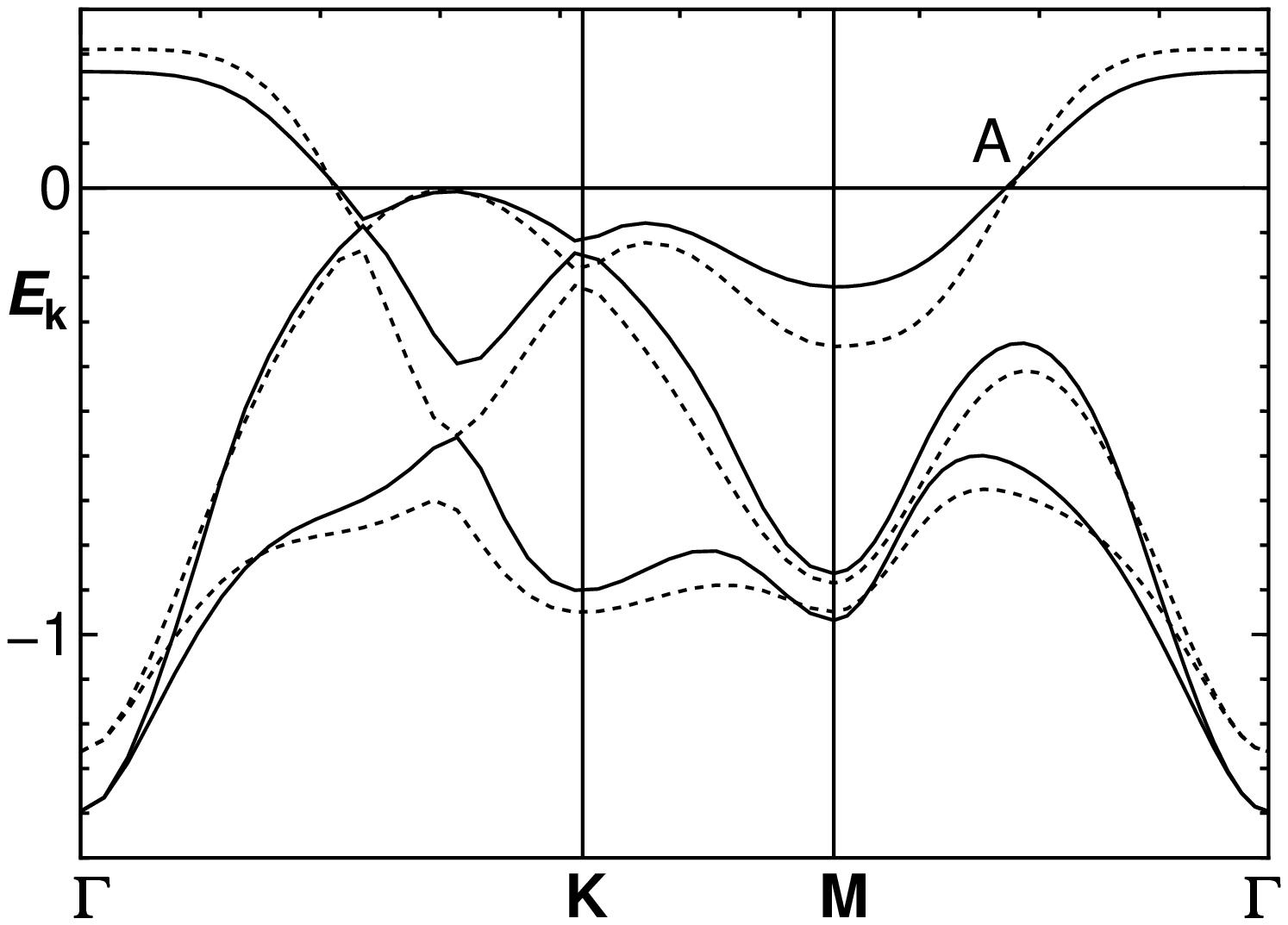}
\end{figure}
\begin{figure}
  \includegraphics[height=.22\textheight]{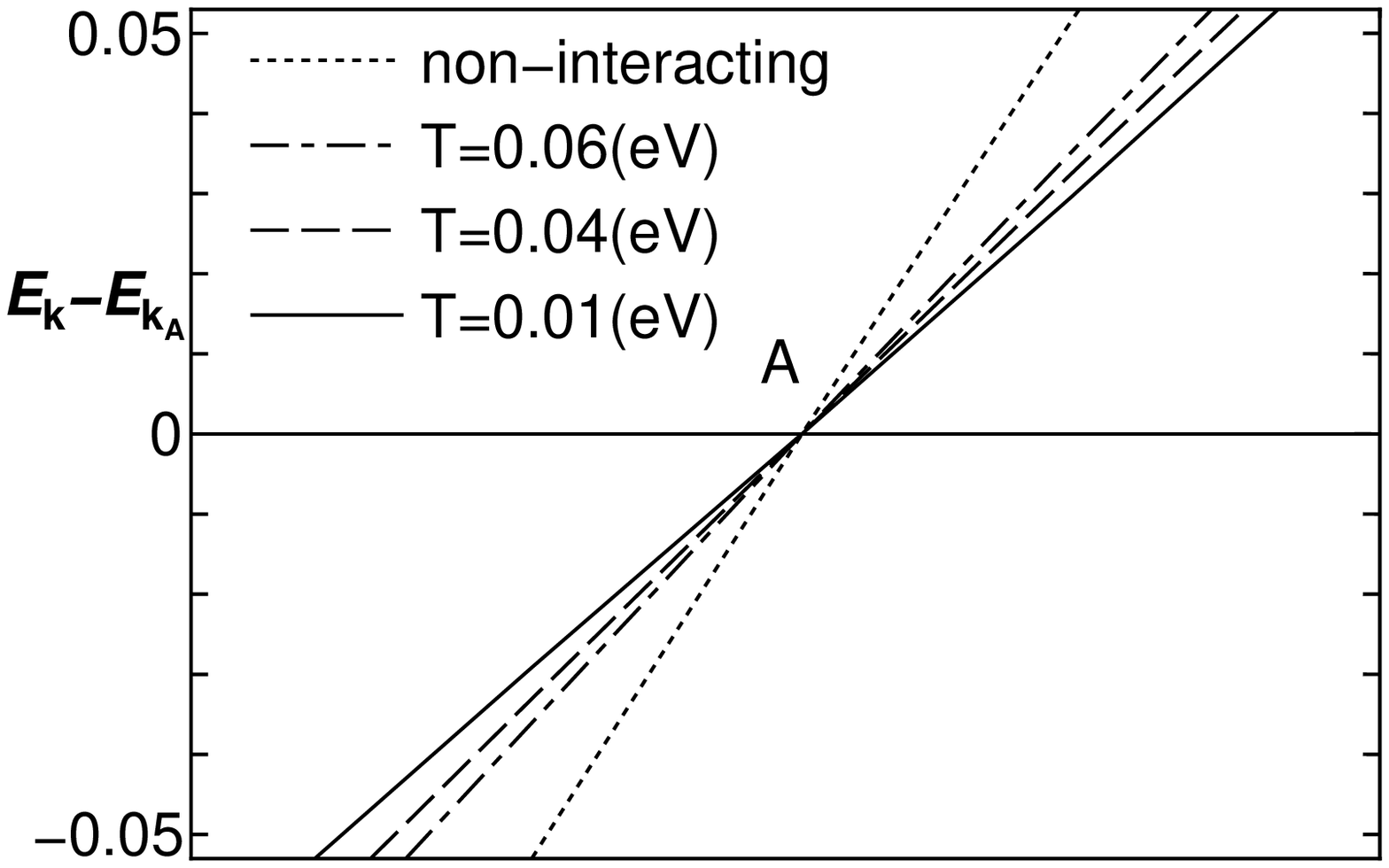}
  \caption{(upper panel) The renormalized band dispersion (solid line)
and the band dispersion without Coulomb interactions (dot line) at $3V_t=0.12$.
(lower panel) Closeup of vicinity of point A at various temperatures.}\label{fig}
\end{figure}
\\We calculate the normal selfenergy $\Sigma({\bf k},\omega)$ in the FLEX approximation scheme.
Then, the renormalized quasiparticle band dispersion $E_{\bf k}$ is decided by the following equation.
\begin{eqnarray*}
\mbox{det}|(E_{\bf k}+\mu){\bf 1}-\hat\epsilon_{\bf k}-\left(\hat\Sigma^R({\bf k},E_{\bf k})+\hat\Sigma^A({\bf k},E_{\bf k})\right)/2|=0.
\end{eqnarray*}
In this case, we take care that both $\hat\epsilon_{\bf k}$ and
$\left(\hat\Sigma^R({\bf k},E_{\bf k})+\hat\Sigma^A({\bf k},E_{\bf k})\right)/2$ are complex Hermitian matrices.
Here we employed the mean value of retarded and advanced selfenergy on the selfenergy
to cancel out the effect of damping of quasiparticle.
By diagonalizing the Green function $\hat G$, we can switch to band-representation.
In the band-representation, mass enhancement factor $m^*/m$ is determined by following equation.

\begin{eqnarray*}
\frac{m^*_\alpha}{m_\alpha}&=&\frac{d \epsilon^\alpha_{\bf k}}{d k}\Big/\frac{d E^\alpha_{\bf k}}{d k}\\
&=&\left(1+\frac{\partial \mbox{Re}\Sigma^\alpha({\bf k},E^\alpha_{\bf k})}{\partial k}\Big/\frac{d\epsilon^\alpha_{\bf k}}{d k}\right)^{-1}\\
&&{}\times\left(1-\frac{\partial \mbox{Re}\Sigma^\alpha({\bf k},E^\alpha_{\bf k})}{\partial E^\alpha_{\bf k}}\right),
\end{eqnarray*}
where $k={\bf k\cdot n_\bot}$ (${\bf n_\bot}$ is a unit vector perpendicular to the FS),
and $\alpha$ indicates the band index.
The first bracket in this equation is so-called $k$-mass and the second bracket is so-called $\omega$-mass.
In strongly correlated electron systems,
$\omega$-mass constitutes the main part of mass enhancement and it makes the effective mass heavy.

In Fig. \ref{fig}, we show the calculated band dispersion of Na$_{0.35}$CoO$_2$.
Near the Fermi level, we notice that the electronic band is renormalized
and the slope of band dispersion decreases at lower temperatures.
The mass enhancement factor at point A in Fig. \ref{fig} are 1.78, 1.73, 1.60 and 1.47 at $T=0.01, 0,02, 0.04$ and $0.06$ (eV) respectively.
The mass enhancement factor would increase further if one take large $U, U'$.
In the three bands we show in Fig. \ref{fig},
the top band which across the Fermi enregy is strongly renormalized in whole.
The other two bands below the Fermi energy are also renormalized for $|E^\alpha_{\bf k}-\mu|\lsim 1$ (eV).
However, $|E^\alpha_{\bf k}-\mu|$ are greater than $|\epsilon^\alpha_{\bf k}-\mu|$ for $|E^\alpha_{\bf k}-\mu|\gsim 1$ (eV),
which is inconsistent with ARPES measurements.
We conclude that the electronic structure for $|E^\alpha_{\bf k}-\mu|\lsim 1$ (eV) is well reproduced.
Higher energy spectrum by the FLEX approximation will be improved if one take the vertex corrections.
The FLEX approximation gives reliable results for lower energy scale.

In summary, we analyzed the multi-orbital Hubbard model for Na$_{0.35}$CoO$_2$ using the FLEX approximation,
and we found the weak pseudgap behavior in the DOS and spin susceptibility when small hole pockets are absent.
The electronic band is renormalized by the electronic correlation,
and mass enhancement factor $m^*/m\simeq 1.5\sim 1.8$.
This result supports the ARPES measurements.

\end{document}